\documentclass{article}

\usepackage[final]{neurips_2026}
  \makeatletter
  \renewcommand{\@noticestring}{%
    Accepted by the 9th Workshop on Machine Learning and the Physical Sciences (ML4PS 2026). Do not distribute.%
  }
  \makeatother

\usepackage[utf8]{inputenc} % allow utf-8 input
\usepackage[T1]{fontenc}    % use 8-bit T1 fonts
\usepackage{hyperref}       % hyperlinks
\usepackage{url}            % simple URL typesetting
\usepackage{booktabs}       % professional-quality tables
\usepackage{amsfonts}       % blackboard math symbols
\usepackage{nicefrac}       % compact symbols for 1/2, etc.
\usepackage{microtype}      % microtypography
\usepackage{xcolor}   % colors
\usepackage{graphicx}
\usepackage{amsmath}
\usepackage{adjustbox}

\title{Leveraging Industrial Foundation Models at the Edge of Particle Physics Detectors via Distillation Learning and Hardware Co-design}

\author{%
  Gia Ancone \\
  Department of Physics \\
  Stanford University \\
  450 Jane Stanford Way, Stanford, CA 94305, United States of America \\
  \texttt{gancone@stanford.edu} \\
  \And
  Qibin Liu \\
  SLAC National Accelerator Laboratory\\
  2575 Sand Hill Road, Menlo Park, CA 94025, United States of America\\
  \texttt{qibin@slac.stanford.edu} \\
  \And
  Liangyu Wu \\
  Department of Physics\\
  Stanford University\\
  450 Jane Stanford Way, Stanford, CA 94305, United States of America \\
  \texttt{liangyu.wu@stanford.edu} \\
  \AND
  Julia Gonski \\
  SLAC National Accelerator Laboratory\\
  2575 Sand Hill Road, Menlo Park, CA 94025, United States of America\\
  \texttt{jgonski@slac.stanford.edu} \\
}

\begin{document}

\maketitle

\begin{abstract}
Data acquisition (DAQ) systems at future particle physics experiments stand to benefit from the extremes of AI/ML development: large-scale foundation models can enhance the performance of feature extraction algorithms, and small-scale on-detector deployments can enable real-time intelligent data handling. 
This work provides the first tokenizer fine-tuning of an industrial foundation model for particle physics DAQ.
Starting from the backbone of Google Research's TimesFM (Time Series Foundation Model), we address real-time regression tasks for drift chamber trackers and dual-readout calorimeters by jointly training the pretrained tokenizer and a newly initialized residual adapter, then fitting task-specific linear heads while keeping the transformer backbone frozen.
Furthermore, the resulting TimesFM teachers can be distilled into students and co-designed with FPGA implementation to support real-time operation at future colliders.
The distilled and compressed models achieve competitive performance, while distillation reduces reliance on labeled training data.
Further, the pipeline of distillation and model compression from TimesFM is generic and can be easily adapted to a variety of 1D waveform tasks across domains. 
\end{abstract}

%%%%%%%%%%%%%%%%%%%%%%%%%%%%%%%%%%%%%%%%%%%%%%%%%%%%%%%%%%%%
\section{Introduction}

%- P1: motivate future experiments, big data frontier opens up door to benefits of neural scaling laws from super large foundation models, but also presents challenges of ultra-high data rate management. 

The future of high energy physics (HEP) is predicated on the ability to design and operate high precision and environmentally tolerant detectors that produce and record a new frontier in big data~\cite{p5_2023,CERN-ESU-2025-002}. 
These high data volumes in turn usher in next-generation artificial intelligence and machine learning (AI/ML) applications for searches, precision measurements, and future design/operation techniques. 
Recent work in high energy collider physics has demonstrated neural scaling laws in the context of jet tagging~\cite{vigl2026neuralscalinglawsboosted}, as well as evidence for transfer learning in large-scale foundation model pre-training to both in~\cite{PhysRevD.111.054015,Bhimji_2026,Mikuni_2025,Birk_2024,hsu2026evenetfoundationmodelparticle} and out~\cite{mikuni2025omnicosmostransferringparticlephysics,wu2026lightweightfoundationmodelcollider} of domain tasks.

%- P2: edge ML for HEP
Harnessing the power of large-scale compute and training datasets for HEP is only feasible if the very large data volumes produced by modern experiments can be appropriately handled and recorded to maintain all key physics information. 
This requires online data handling and management through real-time trigger and data acquisition (TDAQ) systems, which must reconstruct, filter, and transmit events off-detector with $\mathcal{O}$(ns--$\mu$s) timescales~\cite{Aad_2024,Hayrapetyan_2024}. 
Looking towards future colliders, the co-design of such algorithms alongside efficient hardware platforms that can withstand the extreme spatial constraints of the detector environment opens the door to data reduction at the ``edge'', or source of data. 
Previous works in this area have studied lightweight feature extraction or filtering algorithms for future pixel detectors~\cite{yoo2023_on_sensor_filtering}, drift chamber trackers~\cite{yilmaz2025edgemachinelearningcluster}, and dual-readout calorimeters~\cite{wu2026machinelearningenablesrealtime}.

%- P3: this work & lit review 
This work explores the intersection of the extremes of AI/ML for HEP by proposing to improve edge ML for data reduction through \textbf{PRetrained Industrial Models for the Edge of Detectors (PRIMED)}.%
  \footnote{Code Repository: \url{https://anonymous.4open.science/r/p7260/}}
PRIMED uses Google Research's TimesFM (Time Series Foundation Model)~\cite{das2024decoderonlyfoundationmodeltimeseries}, an open-source foundation model trained on hundreds of billions of points from 1D time series datasets, to demonstrate that multi-disciplinary, large-scale pretraining on generic 1D temporal signals provides strong, transferable representations for detector waveforms.

Distilling TimesFM with a task-adapted tokenizer into a student architecture tailored for field programmable gate array (FPGA) implementation, along with subsequent model compression techniques, yields models with competitive performance and reduced reliance on labeled training data for two proposed collider detectors, with synthesis estimates supporting potential real-time front-end implementation.  These results represent first demonstrations of generalizable techniques, namely a) using industrial foundation models for HEP TDAQ tasks and b) distilling/compressing foundation models for edge implementation, with wide-ranging impacts for real-time data acquisition across fundamental physics and beyond.

%%%%%%%%%%%%%%%%%%%%%%%%%%%%%%%%%%%%%%%%%%%%%%%%%%%%%%%%%%%%
\section{Methods}

This work considers two feature extraction tasks for future collider detector subsystems that record 1D waveforms, both burdened by the need for high granularity and high data rate within stringent spatial and power constraints.
The first is particle identification in drift chambers (DCH) by using ML to regress cluster count~\cite{Rolandi:2008ujz}, with performance evaluated by kaon–pion separation $S = \frac{|\frac{dN}{dx}_\pi - \frac{dN}{dx}_K|}{\frac{\sigma_\pi + \sigma_K}{2}}$~\cite{Tian_2025}.
% CHANGED: crystal named; regression targets stated (photon numbers and t0); q defined; timing metric added.
The second concerns dual-readout (DRO) calorimetry~\cite{RevModPhys.90.025002,hirosky2024dualreadoutcalorimetryhomogeneouscrystals,refId0,Eno:2025ltc} with a PbWO$_4$ (PWO) crystal and ML-based regression of the waveform's Cherenkov (C) and scintillation (S) photon numbers and of the signal time offset $t_0$, with performance quantified by the 68\% containment error $\mathrm{err68} = P_{68}\left(\frac{|\hat{q} - q|}{q} \times 100\%\right)$ on the reconstructed Cherenkov-to-scintillation ratio $q = C/S$ (for events where $C > 0$ and $S > 0$), and by the 68th percentile of $|\hat{t}_0 - t_0|$ for the timing.
These two tasks represent different kinds of challenges: the DCH task is inherently local; accurate cluster counting relies on point-by-point information, as even a single waveform point can indicate a primary cluster.
% CHANGED: "template fractions" -> "components" (the networks regress photon numbers, not template fractions).
By contrast, the DRO task is more global, requiring knowledge of the overall waveform shape to extract meaningful C and S components.

The DCH simulated dataset~\cite{Tian_2025,cepc_dch_data} comprises 500,000 waveforms of incident charged pions ($\pi^\pm$) with momenta of 0.2--20\,GeV/$c$ for training, and a separate 10\,GeV/$c$ production of $\pi^\pm$ and charged kaons ($K^\pm$) for evaluation.
% CHANGED: the PWO study does not use the e- / K0L samples; it uses 100,000 simulated SiPM waveforms with random C and S.
The DRO simulated dataset is based on the waveform simulation of~\cite{wu2026machinelearningenablesrealtime}, modified to make the task more realistic: the numbers of Cherenkov and scintillation photons are drawn independently and uniformly from 0--1500 and 0--900, so that the total signal alone does not uniquely determine their ratio, $t_0$ is randomized within $\pm2.5$\,ns, and silicon-photomultiplier dark counts (10\,MHz), single-photoelectron amplitude spread (10\%), optical crosstalk (7\%) and electronic noise are added. It comprises 100,000 PWO waveforms of 6246 samples at 0.32\,ns, of which 80,000 form the training and validation pool and 20,000 are held out for evaluation.
The target platform is an FPGA, with a potential path toward radiation-hard ASIC integration through embedded FPGA (eFPGA) technology~\cite{Gonski_2024}.

The PRIMED pipeline is as follows: 1. train the pretrained TimesFM tokenizer and a newly initialized residual adapter while keeping the transformer backbone frozen, then fit a task-specific ridge regression head; 2. distill into a student ML architecture that comprises only layers which can be synthesized for FPGA implementation; 3. leverage hardware compression techniques (pruning and quantization) to enable final models that respect operational constraints. 

%-------------------------------------------
\subsection{TimesFM Tokenizer Adaptation}

Both teachers start from the public TimesFM 2.5 checkpoint
(\texttt{timesfm-2.5-200m-pytorch}, 231M parameters), which patches a waveform into 32-sample tokens of dimension 1280. We discard the forecasting output and train the pretrained tokenizer together with a newly initialized residual convolutional adapter, keeping all transformer blocks frozen. The adapter adds a residual correction to the tokenizer embeddings and is initialized to produce zero correction. Self-supervised training combines next-patch prediction and masked-patch reconstruction, using two temporary linear reconstruction heads. These heads provide the tokenizer-training signal and are not used for downstream regression.
After self-supervised training, we freeze the tokenizer and adapter and fit linear ridge regression heads on mean-pooled token embeddings, predicting the cluster count for DCH, and $C$, $S$, and $t_0$ for DRO. The ridge penalty strength is selected to minimize validation error.
To assess the contribution of pretraining, we repeat tokenizer adaptation and ridge-head fitting with randomly initialized, frozen transformer blocks.

For both tasks, tokenizer adaptation uses AdamW~\cite{loshchilov2019decoupledweightdecayregularization} with a cosine schedule, weight decay of 0.01, gradient clipping of 1.0, $\mathrm{lr}=10^{-4}$ for the tokenizer and adapter and $\mathrm{lr}=10^{-3}$ for the reconstruction heads and mask token, batch size 16, and a masking probability of 25\%.
For DCH, adaptation uses the full 500,000-waveform training pool without labels for three epochs, totaling 93,750 optimizer steps. The resulting tokenizer is shared across all four label fractions and three downstream seeds.
For DRO, adaptation uses the full 80,000-waveform development pool without labels for three epochs, totaling 15,000 optimizer steps. A separate tokenizer is trained for each seed and shared across all five label fractions.

The raw DCH and DRO waveforms contain 3,000 and 6,246 samples, respectively, and are edge-padded to 3,008 and 6,272 samples.
DCH uses per-waveform mean/std normalization followed by cumulative patch-wise RevIN normalization, whereas DRO uses global min-max scaling with constants from its tokenizer-training pool.
Each task uses the same preprocessing during training and evaluation.
% TODO: Confirm that the new PWO teacher retains the reference pipeline's normalization.
At the full label fraction, DCH regression heads are fitted on 450,000 events, validated on 50,000, and evaluated on 10\,GeV/$c$ samples of 100,000 pions and 100,000 kaons. For DRO, regression heads are fitted on 72,000 events, validated on 8,000, and evaluated on an independent 20,000-event test set. Both DCH and DRO runs used NVIDIA A100 (80 GB) GPUs at the National Energy Research Scientific Computing Center (NERSC).
% Add the separate 500,000-event DCH regression test pool where count MAE/RMSE
% results are described. It is not the species sample used for separation.
% Restore a combined GPU-hour total only after updated job accounting.

%-------------------------------------------
\subsection{Distillation Learning}
Both students are fully connected networks with hidden widths 16--24--8, following the published dual-readout edge model \cite{wu2026machinelearningenablesrealtime}. They differ only in input/output dimensions: DCH uses 602 inputs and one output (10,265 parameters), while DRO uses 640 inputs and three outputs (10,891 parameters). Inputs are stride-downsampled waveforms --- $3{,}008\to602$ at stride 5 for DCH and $6{,}246\to625$ at stride 10 for DRO, with the latter edge-padded to 640 --- then globally min-max scaled using the DCH student-training rows or the DRO development pool. The DCH student is randomly initialized, with its output bias set to the mean scaled training target.

Distillation targets come from a single teacher checkpoint per fraction and seed. For both tasks, the matched teacher's predictions are cached and used as targets for all student-training rows, including those with available true labels, at all label fractions. Students minimize mean squared error against these predictions, with task-specific target scaling and loss weights. Validation rows are excluded from training, and their true labels are used for checkpoint selection. For DCH, targets are divided by 53, the MSE is weighted by $53^2/100000$, and predictions are multiplied by 53 for evaluation. For DRO, targets are standardized using means $[750,450,0]$ and scales $[433,260,1.44]$ for $(C,S,t_0)$, with loss weights $[1,1,0.01]$; predictions are transformed back to physical units for evaluation.

For each task, the from-scratch control matches the distilled student in architecture, input preprocessing, optimizer settings, maximum training budget, and stopping rule, but is trained only on true labels from the same labeled training subset used to fit the teacher. DCH uses Adam~\cite{kingma2017adammethodstochasticoptimization} with an initial $\mathrm{lr}=10^{-3}$, decaying to $10^{-5}$, no weight decay, and batch size 64. Training runs for up to 600,000 steps at every fraction, with validation-based early stopping and checkpoint selection. DRO uses Adam with $\mathrm{lr}=10^{-2}$, 5\% linear warm-up followed by cosine decay, batch size 1024, and 800,000 steps, selecting the checkpoint with the lowest smoothed validation loss.
% TODO: Confirm that these DRO settings apply to all plotted fractions and seeds
% and to the matched from-scratch controls.

%-------------------------------------------
\subsection{Hardware Co-Design}

Models are synthesized from software descriptions to FPGA implementations using \texttt{hls4ml}~\cite{fastml_hls4ml, Duarte:2018ite}, which also provides estimates of the FPGA resources in look-up tables (LUTs), flip-flops (FFs), digital signal processors (DSPs), and operational latency.
Both designs were synthesized targeting the AMD Xilinx XCVU13P-FLGA2577-2-E FPGA with a 25 ns clock period (40 MHz) and ReuseFactor = 1, using a latency-oriented scheduling strategy.
These figures reflect C-synthesis estimates only and exclude waveform acquisition overheads and system-level I/O effects.
To further compress the student model for hardware acceleration, models were subject to pruning to reduce the computational burden of low-magnitude weights, and quantization-aware training (QAT) via \texttt{QKeras}~\cite{qkeras} to reduce the size of the representation of models' weights/biases as well as training/inference values.

Both models are pruned and then quantization-aware trained before synthesis; for DRO, the float model is first rescaled per unit and fine-tuned with bounded weights.
In pruning~\cite{han2015pruning,tfmot}, a
fraction of the smallest-magnitude kernel weights is set to zero (50\% for DCH; 70\% of the input layer for DRO, giving 81\% zero weights once inactive units are removed), without changing the
network architecture. The model is fine-tuned for up to 30 (DCH) or 200 (DRO) epochs while the pruning mask is updated, after which the mask is fixed and QAT~\cite{jacob2018qat,coelho2021qkeras} retrains the remaining weights at the target precision, so the network learns to compensate for the rounding error. DCH uses 10-bit power-of-two weights in every layer; DRO uses 10-bit power-of-two weights only in the input layer, where most weights sit, and 10-bit fixed-point weights on the smaller layers, with integer bits $i = 3, 3, 2$. DCH biases and outputs use signed $\langle 12,3\rangle$ precision and hidden activations unsigned $\langle 12,2\rangle$; DRO uses signed $\langle 10,3\rangle$ and unsigned $\langle 10,2\rangle$, respectively.
Power-of-two multiplications are implemented as bit shifts, and the synthesized designs use LUT logic for the remaining multiplications, yielding DSP = 0 for both designs.

Both compressed models (Table~\ref{tab:results}) achieve 25\,ns latency with an initiation interval of 1 and use no DSPs, supporting lightweight real-time FPGA operation.
LUT usage is on par with published benchmarks~\cite{yilmaz2025edgemachinelearningcluster,wu2026machinelearningenablesrealtime}.

\begin{table}[h]
\centering
{\small
\setlength{\tabcolsep}{4pt}
\begin{tabular}{c|ccccccc}
\toprule
Method & Dense Params & Sparsity & Quantization & LUTs & FFs & DSPs & Latency [ns] \\
\midrule
DCH & 10{,}265 & 50\% & $\langle 10,\mathrm{po2}\rangle$ kernels & 226{,}070 & 14{,}608 & 0 & 25 \\
DRO & 10{,}891 & 81\% & $\langle 10,\mathrm{po2}\rangle$ / $\langle 10,i\rangle$ kernels & 96{,}714 & 15{,}424 & 0 & 25 \\
\bottomrule
\end{tabular}
}

\caption{Model compression approach (dense parameter count and sparsity), FPGA resources (LUTs, FFs, and DSPs), and C-synthesis latency estimates for the DCH and DRO tasks.
Sparsity is the fraction of zero kernel weights.
$\langle b,\mathrm{po2}\rangle$ denotes power-of-two quantization with $b$ total bits.
$\langle b,i\rangle$ denotes fixed-point quantization with $b$ total bits, including $i$ integer bits (sign bit included for signed values).
DCH kernels are power-of-two in all layers; biases and outputs use signed $\langle 12,3\rangle$, hidden activations unsigned $\langle 12,2\rangle$, and inputs unsigned $\langle 24,1\rangle$ precision.
DRO kernels are power-of-two in the first layer, which holds 94\% of the weights, and fixed point in the remaining layers ($i = 3, 3, 2$); biases and outputs use signed $\langle 10,3\rangle$, hidden activations unsigned $\langle 10,2\rangle$, and inputs signed $\langle 24,1\rangle$ precision.
Both designs target the XCVU13P at a 25\,ns clock with reuse factor 1 and LUT-based multiplication, and complete in one clock cycle.}
\label{tab:results}
\end{table}

%%%%%%%%%%%%%%%%%%%%%%%%%%%%%%%%%%%%%%%%%%%%%%%%%%%%%%%%%%%%
\section{Results}

Two performance criteria are considered when evaluating the utility of PRIMED for edge DAQ tasks. The first is whether the resulting distilled and compressed model achieves competitive performance relative to task-specific ML models while meeting edge deployment constraints. The second is whether the distilled student reduces reliance on labeled data, namely whether it can match or outperform a model trained from scratch with the same architecture using fewer labeled training instances.

\begin{figure}[!htbp]
\centering
\includegraphics[width=0.49\textwidth]{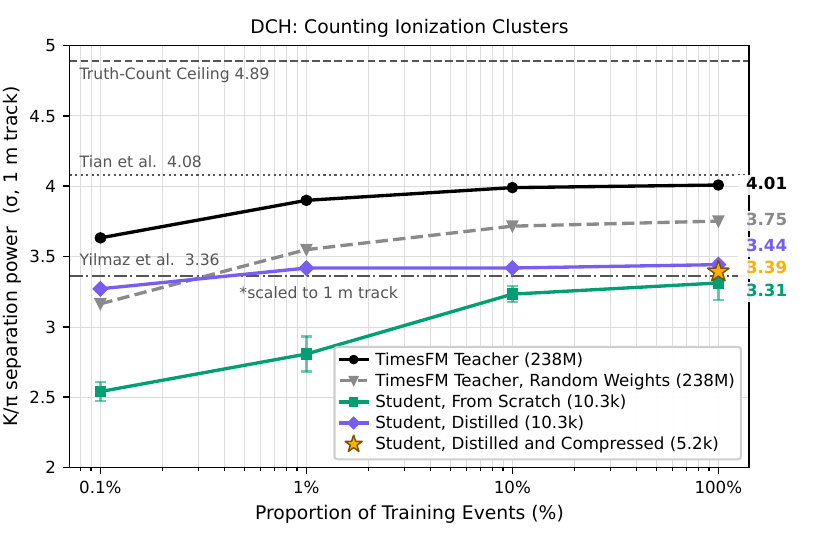}
\includegraphics[width=0.49\textwidth]{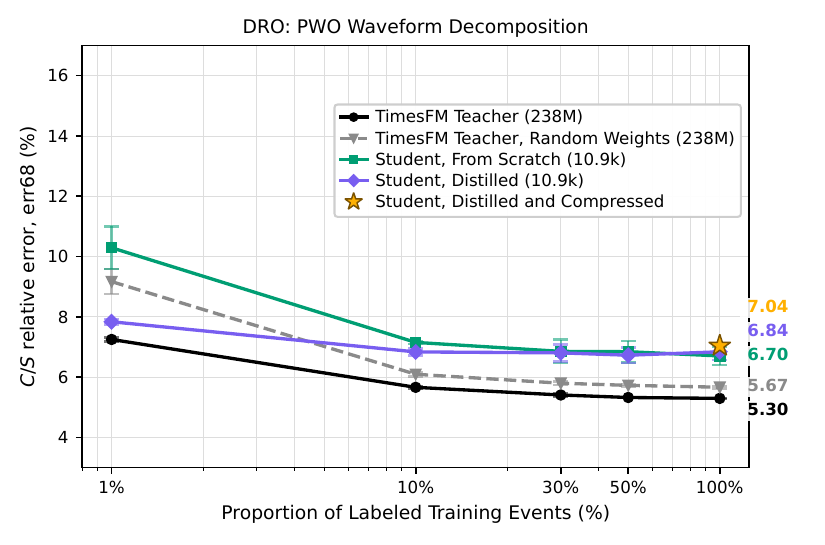}
\caption{PRIMED performance versus labeled-data fraction for DCH cluster counting (left, separation $S$) and DRO waveform decomposition on PWO data (right, $\mathrm{err68}$ on $C/S$). Curves show the tokenizer-adapted TimesFM teacher, its random frozen-transformer counterpart, the from-scratch student, and the student distilled from the pretrained teacher. Stars mark compressed distilled students at 100\% labeled data. DCH horizontal references are from Refs.~\cite{Tian_2025,yilmaz2025edgemachinelearningcluster}. The edge-DNN reference, $3.36\sigma$, scales the reproduced 10\,GeV result of $4.75\sigma$ from 2\,m to 1\,m using square-root track-length scaling. Curves show three-seed means $\pm$ one standard deviation, except the DCH random-transformer teacher and distilled student, which use seed 0.}
\label{fig:fmplots}
\end{figure}

Figure~\ref{fig:fmplots} compares predictive performance and label efficiency for both tasks, including the compressed models in Table~\ref{tab:results}.
Three key observations can be made. First, the final PRIMED output models achieve competitive performance after compression, with the DCH model exceeding the published edge-DNN benchmark for a model of comparable size (Table~\ref{tab:results}). Second, the distilled student has favorable data scaling properties, outperforming its from-scratch counterpart at all evaluated fractions for DCH and achieving lower mean $C/S$ error at all fractions below 100\% for DRO; the DCH distilled student trained on 1\% of the labeled data exceeds the mean performance of the from-scratch student trained on the full dataset. For DRO, distillation reduces error by approximately 24\% at 1\% labeled data and 4.4\% at 10\%. This last conclusion is especially relevant for development of edge ML for future experiments, where high volumes of labeled training data are not available at the time of detector design or commissioning. Third, for both tasks, the pretrained teacher outperforms its random-weights counterpart, supporting the transfer learning benefit of the pretrained backbone beyond task-specific tokenizer adaptation alone. Hence, we see the benefit of distilling from a pretrained foundation model for HEP TDAQ tasks.

While these results are a valuable first proof-of-concept for the PRIMED pipeline, they remain limited in scope. This work only considers one student architecture, distillation technique, and hardware synthesis approach; gains in performance could be achieved from optimizing these steps. A next step is to train a unified physics waveform tokenizer across detector subsystems and evaluate its transfer to new tasks. Additional 1D physics waveform datasets would help test whether these transfer learning benefits extend across physics tasks.

The ultimate front-end feasibility of the algorithms described here will depend on the resources available in a target experimental setup, particularly the power budget. Some front-end applications can make use of commercial FPGAs like the device used here, while others will require customized hardware such as eFPGA, where the available logic depends on the number of FPGA tiles integrated into the ASIC design. These system aspects, including power consumption, are left to later work.  Additionally, as these results are simulation-only, replication in test beam data is essential before edge ML feature compression can be seriously considered in future detector design.

%%%%%%%%%%%%%%%%%%%%%%%%%%%%%%%%%%%%%%%%%%%%%%%%%%%%%%%%%%%
\begin{ack}
This work is supported by the U.S. Department of Energy under contract number DE-AC02-76SF00515 and the Office of the Vice Provost for Undergraduate Education at Stanford University. This research used resources of the National Energy Research
Scientific Computing Center, a DOE Office of Science User Facility
supported by the Office of Science of the U.S. Department of Energy
under Contract No. DE-AC02-05CH11231 using NERSC award
HEP-ERCAP0037461.
\end{ack}

\bibliographystyle{unsrt} 
\bibliography{fm-real-time}

@misc{das2024decoderonlyfoundationmodeltimeseries,
      title={A decoder-only foundation model for time-series forecasting}, 
      author={Abhimanyu Das and Weihao Kong and Rajat Sen and Yichen Zhou},
      year={2024},
      eprint={2310.10688},
      archivePrefix={arXiv},
      primaryClass={cs.CL},
      url={https://arxiv.org/abs/2310.10688}, 
       note={Code and model weights released under Apache-2.0}

}

@software{tfmot,
    title     = {TensorFlow Model Optimization Toolkit},
    author    = {{TensorFlow Authors}},
    year      = {2023},
    publisher = {GitHub},
    url       = {https://github.com/tensorflow/model-optimization},
    note      = {Apache-2.0}
}

@inproceedings{yilmaz2025edgemachinelearningcluster,
  title     = {Edge Machine Learning for Cluster Counting in Next-Generation Drift Chambers},
  author    = {Deniz Yilmaz and Liangyu Wu and Julia Gonski and Dylan Rankin and Christian Herwig},
  booktitle = {Machine Learning and the Physical Sciences Workshop, NeurIPS 2025},
  year      = {2025},
  eprint    = {2511.10540},
  archivePrefix = {arXiv},
  primaryClass  = {physics.ins-det},
  url       = {https://arxiv.org/abs/2511.10540}
}

@misc{wu2026machinelearningenablesrealtime,
      title={Machine Learning Enables Real-Time Waveform Decomposition for Dual-Readout Calorimetry}, 
      author={Liangyu Wu and Qibin Liu and Marco Toliman Lucchini and Julia Gonski and Marcello Campajola and Stefano Moneta},
      year={2026},
      eprint={2604.26090},
      archivePrefix={arXiv},
      primaryClass={physics.ins-det},
      url={https://arxiv.org/abs/2604.26090}, 
}

@article{Duarte:2018ite,
    author = "Duarte, Javier and others",
    title = "{Fast inference of deep neural networks in FPGAs for particle physics}",
    eprint = "1804.06913",
    archivePrefix = "arXiv",
    primaryClass = "physics.ins-det",
    reportNumber = "FERMILAB-PUB-18-089-E",
    doi = "10.1088/1748-0221/13/07/P07027",
    journal = "JINST",
    volume = "13",
    number = "07",
    pages = "P07027",
    year = "2018"
}

@software{fastml_hls4ml,
  author       = {{FastML Team}},
  title        = {fastmachinelearning/hls4ml},
  year         = 2026,
  publisher    = {Zenodo},
  version      = {v1.3.0},
  doi          = {10.5281/zenodo.1201549},
  url          = {https://github.com/fastmachinelearning/hls4ml},
  note         = {Apache-2.0}

}

@misc{p5_2023,
  title={{Exploring the Quantum Universe: Pathways to Innovation and Discovery in Particle Physics}},
  author={{H. Murayama, K. Heeger, et al}},
  year = {2023},
  month = {Dec},
  note = {https://www.usparticlephysics.org/2023-p5-report/}
}

@techreport{CERN-ESU-2025-002,
      title         = "{The European Strategy for Particle Physics: 2026 Update -
                       Recommendations by the European Strategy Group}",
      reportNumber  = "CERN-ESU-2025-002",
      address       = "Geneva",
      year          = "2025",
      url           = "https://cds.cern.ch/record/2950671",
      doi           = "10.17181/CERN.423R.S20Z",
}

@misc{vigl2026neuralscalinglawsboosted,
      title={Neural Scaling Laws for Boosted Jet Tagging}, 
      author={Matthias Vigl and Nicole Hartman and Michael Kagan and Lukas Heinrich},
      year={2026},
      eprint={2602.15781},
      archivePrefix={arXiv},
      primaryClass={hep-ex},
      url={https://arxiv.org/abs/2602.15781}, 
}

@misc{wu2026lightweightfoundationmodelcollider,
      title={A Lightweight Foundation Model for Collider Physics with Multi-Domain Adaptation}, 
      author={Liangyu Wu and Qibin Liu and Alexander Yue and Julia Gonski},
      year={2026},
      eprint={2607.27501},
      archivePrefix={arXiv},
      primaryClass={cs.LG},
      url={https://arxiv.org/abs/2607.27501}, 
}

@misc{hsu2026evenetfoundationmodelparticle,
      title={EveNet: A Foundation Model for Particle Collision Data Analysis}, 
      author={Ting-Hsiang Hsu and Bai-Hong Zhou and Qibin Liu and Yue Xu and Shu Li and George Wei-Shu Hou and Benjamin Nachman and Shih-Chieh Hsu and Vinicius Mikuni and Yuan-Tang Chou and Yulei Zhang},
      year={2026},
      eprint={2601.17126},
      archivePrefix={arXiv},
      primaryClass={hep-ex},
      url={https://arxiv.org/abs/2601.17126}, 
}

@misc{mikuni2025omnicosmostransferringparticlephysics,
      title={OmniCosmos: Transferring Particle Physics Knowledge Across the Cosmos}, 
      author={Vinicius Mikuni and Ibrahim Elsharkawy and Benjamin Nachman},
      year={2025},
      eprint={2512.24422},
      archivePrefix={arXiv},
      primaryClass={astro-ph.CO},
      url={https://arxiv.org/abs/2512.24422}, 
}

@article{PhysRevD.111.054015,
  title = {Method to simultaneously facilitate all jet physics tasks},
  author = {Mikuni, Vinicius and Nachman, Benjamin},
  journal = {Phys. Rev. D},
  volume = {111},
  issue = {5},
  pages = {054015},
  numpages = {15},
  year = {2025},
  month = {Mar},
  publisher = {American Physical Society},
  doi = {10.1103/PhysRevD.111.054015},
  url = {https://link.aps.org/doi/10.1103/PhysRevD.111.054015}
}

@article{Bhimji_2026,
   title={Foundation model framework for all tasks involving jet physics},
   volume={113},
   ISSN={2470-0029},
   url={http://dx.doi.org/10.1103/knmd-f5jm},
   DOI={10.1103/knmd-f5jm},
   number={3},
   journal={Physical Review D},
   publisher={American Physical Society (APS)},
   author={Bhimji, Wahid and Harris, Chris and Mikuni, Vinicius and Nachman, Benjamin},
   year={2026},
   month=feb }

@article{Mikuni_2025,
   title={Solving key challenges in collider physics with foundation models},
   volume={111},
   ISSN={2470-0029},
   url={http://dx.doi.org/10.1103/PhysRevD.111.L051504},
   DOI={10.1103/physrevd.111.l051504},
   number={5},
   journal={Physical Review D},
   publisher={American Physical Society (APS)},
   author={Mikuni, Vinicius and Nachman, Benjamin},
   year={2025},
   month=mar }

@article{Birk_2024,
   title={{OmniJet-$\alpha$}: the first cross-task foundation model for particle physics},
   volume={5},
   ISSN={2632-2153},
   url={http://dx.doi.org/10.1088/2632-2153/ad66ad},
   DOI={10.1088/2632-2153/ad66ad},
   number={3},
   journal={Machine Learning: Science and Technology},
   publisher={IOP Publishing},
   author={Birk, Joschka and Hallin, Anna and Kasieczka, Gregor},
   year={2024},
   month=Aug, pages={035031} }

@article{Aad_2024,
   title={The ATLAS trigger system for LHC Run 3 and trigger performance in 2022},
   volume={19},
   ISSN={1748-0221},
   url={http://dx.doi.org/10.1088/1748-0221/19/06/P06029},
   DOI={10.1088/1748-0221/19/06/p06029},
   number={06},
   journal={Journal of Instrumentation},
   publisher={IOP Publishing},
   year={2024},
   month=June,
   pages={P06029}
}

@article{Hayrapetyan_2024,
   title={Development of the CMS detector for the CERN LHC Run 3},
   volume={19},
   ISSN={1748-0221},
   url={http://dx.doi.org/10.1088/1748-0221/19/05/P05064},
   DOI={10.1088/1748-0221/19/05/p05064},
   number={05},
   journal={Journal of Instrumentation},
   publisher={IOP Publishing},
   year={2024},
   month=May,
   pages={P05064}
}

@article{yoo2023_on_sensor_filtering,
      title={{Smart pixel sensors: towards on-sensor filtering of pixel clusters with deep learning}}, 
      author={Jieun Yoo and Jennet Dickinson and Morris Swartz and Giuseppe Di Guglielmo and Alice Bean and Douglas Berry and Manuel Blanco Valentin and Karri DiPetrillo and Farah Fahim and Lindsey Gray and James Hirschauer and Shruti R. Kulkarni and Ron Lipton and Petar Maksimovic and Corrinne Mills and Mark S. Neubauer and Benjamin Parpillon and Gauri Pradhan and Chinar Syal and Nhan Tran and Dahai Wen and Aaron Young},
      year={2024},
      journal={Mach. Learn.: Sci. Technol.},
      publisher={IOP Publishing},
      volume={5},
      pages={035047},
      DOI={10.1088/2632-2153/ad6a00},
      url={https://iopscience.iop.org/article/10.1088/2632-2153/ad6a00}
}

@article{Gonski_2024,
doi = {10.1088/1748-0221/19/08/P08023},
url = {https://doi.org/10.1088/1748-0221/19/08/P08023},
year = {2024},
month = {aug},
publisher = {IOP Publishing},
volume = {19},
number = {08},
pages = {P08023},
author = {Gonski, J. and Gupta, A. and Jia, H. and Kim, H. and Rota, L. and Ruckman, L. and Dragone, A. and Herbst, R.},
title = {Embedded FPGA developments in 130 nm and 28 nm CMOS for machine learning in particle detector readout},
journal = {Journal of Instrumentation}
}

@book{Rolandi:2008ujz,
    author = "Rolandi, L. and Riegler, W. and Blum, W.",
    title = "{Particle Detection with Drift Chambers}",
    doi = "10.1007/978-3-540-76684-1",
    isbn = "978-3-540-76683-4, 978-3-642-09538-2, 978-3-540-76684-1",
    publisher = "Springer",
    series = "Particle Acceleration and Detection",
    year = "2008"
}

@article{refId0,
	author = {{Akchurin, Nural} and {Cash, James} and {Damgov, Jordan} and {Delashaw, Xander} and {Lamichhane, Kamal} and {Harris, Miles} and {Kelley, Mitch} and {Kunori, Shuichi} and {Mergate-Cacace, Harold} and {Peltola, Timo} and {Schneider, Odin} and {Sewell, Julian}},
	title = {High-granularity Dual-readout Calorimeter: Evolution of a Classic Prototype},
	DOI= "10.1051/epjconf/202532000028",
	url= "https://doi.org/10.1051/epjconf/202532000028",
	journal = {EPJ Web Conf.},
	year = 2025,
	volume = 320,
	pages = "00028",
}

@inproceedings{hirosky2024dualreadoutcalorimetryhomogeneouscrystals,
  author       = {Hirosky, R. and Anderson, T. and Cummings, G. and Dubnowski, M. and Guinto-Brody, C. and Guo, Y. and Ledovskoy, A. and Levin, D. and Madrid, C. and Martin, C. and Zhu, J.},
  title        = {Dual-readout calorimetry with homogeneous crystals},
  booktitle    = {Proceedings of CALOR2024},
  series       = {EPJ Web of Conferences},
  year         = {2024},
  note         = {arXiv:2408.11973},
  url          = {https://arxiv.org/abs/2408.11973}
}

@article{RevModPhys.90.025002,
  title = {Dual-readout calorimetry},
  author = {Lee, Sehwook and Livan, Michele and Wigmans, Richard},
  journal = {Rev. Mod. Phys.},
  volume = {90},
  issue = {2},
  pages = {025002},
  numpages = {40},
  year = {2018},
  month = {Apr},
  publisher = {American Physical Society},
  doi = {10.1103/RevModPhys.90.025002},
  url = {https://link.aps.org/doi/10.1103/RevModPhys.90.025002}
}

@article{Eno:2025ltc,
    author = "Eno, S. and Wu, L. and Aamir, M. Y. and Chekanov, S. V. and Nabili, S. and Palmer, C.",
    title = "{On the resolution of dual readout calorimeters}",
    eprint = "2501.15329",
    archivePrefix = "arXiv",
    primaryClass = "physics.ins-det",
    doi = "10.1016/j.nima.2025.171080",
    journal = "Nucl. Instrum. Meth. A",
    volume = "1083",
    pages = "171080",
    year = "2026"
}

@article{Tian_2025,
   title={{Cluster counting algorithm for the CEPC drift chamber using LSTM and DGCNN}},
   volume={36},
   ISSN={2210-3147},
   url={http://dx.doi.org/10.1007/s41365-025-01670-y},
   DOI={10.1007/s41365-025-01670-y},
   number={7},
   journal={Nuclear Science and Techniques},
   publisher={Springer Science and Business Media LLC},
   author={Tian, Zhe-Fei and Zhao, Guang and Wu, Ling-Hui and Zhang, Zhen-Yu and Zhou, Xiang and Xin, Shui-Ting and Liu, Shuai-Yi and Li, Gang and Dong, Ming-Yi and Sun, Sheng-Sen},
   year={2025},
   month=may }

@software{qkeras,
  title        = {QKeras: a quantization deep learning library for Tensorflow Keras},
  year         = 2021,
  publisher    = {github},
  version      = {v0.9.0},
  url          = {https://github.com/google/qkeras},
  note         = {Apache-2.0}

}

@inproceedings{han2015pruning,
  title     = {Learning both Weights and Connections for Efficient Neural Networks},
  author    = {Han, Song and Pool, Jeff and Tran, John and Dally, William J.},
  booktitle = {Advances in Neural Information Processing Systems},
  volume    = {28},
  pages     = {1135--1143},
  year      = {2015},
  eprint    = {1506.02626},
  archivePrefix = {arXiv}
}

@inproceedings{jacob2018qat,
  title     = {Quantization and Training of Neural Networks for Efficient Integer-Arithmetic-Only Inference},
  author    = {Jacob, Benoit and Kligys, Skirmantas and Chen, Bo and Zhu, Menglong and Tang, Matthew and Howard, Andrew and Adam, Hartwig and Kalenichenko, Dmitry},
  booktitle = {Proceedings of the IEEE Conference on Computer Vision and Pattern Recognition (CVPR)},
  pages     = {2704--2713},
  year      = {2018},
  doi       = {10.1109/CVPR.2018.00286},
  eprint    = {1712.05877},
  archivePrefix = {arXiv}
}

@article{coelho2021qkeras,
  title   = {Automatic heterogeneous quantization of deep neural networks for low-latency inference on the edge for particle detectors},
  author  = {Coelho Jr., Claudionor N. and Kuusela, Aki and Li, Shan and Zhuang, Hao and Ngadiuba, Jennifer and Aarrestad, Thea Klaeboe and Loncar, Vladimir and Pierini, Maurizio and Pol, Adrian Alan and Summers, Sioni},
  journal = {Nature Machine Intelligence},
  volume  = {3},
  number  = {8},
  pages   = {675--686},
  year    = {2021},
  doi     = {10.1038/s42256-021-00356-5},
  eprint  = {2006.10159},
  archivePrefix = {arXiv}
}

@misc{cepc_dch_data,
    title     = {Cluster Counting Algorithm for Drift Chamber using {LSTM} and {DGCNN}},
    author    = {Tian, Zhefei and Zhao, Guang and Wu, Linghui and Zhang, Zhenyu and
                 Zhou, Xiang and Xin, Shuiting and Liu, Shuaiyi and Li, Gang and
                 Dong, Mingyi and Sun, Shengsen},
    year      = {2024},
    publisher = {Science Data Bank},
    version   = {V2},
    doi       = {10.57760/sciencedb.16322},
    url       = {https://doi.org/10.57760/sciencedb.16322},
    note      = {Dataset, CC BY-NC-ND 4.0}
  }

@misc{loshchilov2019decoupledweightdecayregularization,
      title={Decoupled Weight Decay Regularization}, 
      author={Ilya Loshchilov and Frank Hutter},
      year={2019},
      eprint={1711.05101},
      archivePrefix={arXiv},
      primaryClass={cs.LG},
      url={https://arxiv.org/abs/1711.05101}, 
}

@misc{kingma2017adammethodstochasticoptimization,
      title={Adam: A Method for Stochastic Optimization}, 
      author={Diederik P. Kingma and Jimmy Ba},
      year={2017},
      eprint={1412.6980},
      archivePrefix={arXiv},
      primaryClass={cs.LG},
      url={https://arxiv.org/abs/1412.6980}, 
}

%%%%%%%%%%%%%%%%%%%%%%%%%%%%%%%%%%%%%%%%%%%%%%%%%%%%%%%%%%%%
%\appendix
%%%%%%%%%%%%%%%%%%%%%%%%%%%%%%%%%%%%%%%%%%%%%%%%%%%%%%%%%%%%

% \newpage
% \input{checklist.tex}

\end{document}